\documentclass[iop,revtex4]{emulateapj}
\usepackage{lineno}

\begin{document}

\renewcommand{\topfraction}{1.0}
\renewcommand{\bottomfraction}{1.0}
\renewcommand{\textfraction}{0.0}

\title{Speckle interferometry at SOAR in 2015\altaffilmark{\dag} }

\altaffiltext{\dag}{Based on observations obtained  at the Southern Astrophysical Research
(SOAR) telescope. }

\author{Andrei Tokovinin}
\affil{Cerro Tololo Inter-American Observatory, Casilla 603, La Serena, Chile}
\email{atokovinin@ctio.noao.edu}
\author{Brian D. Mason \& William I. Hartkopf}
\affil{U.S. Naval Observatory, 3450 Massachusetts Ave., Washington, DC, USA}
\email{bdm@usno.navy.mil, wih@usno.navy.mil}
\author{Rene A. Mendez}
\affil{Universidad de Chile,  Casilla 36-D, Santiago, Chile}
\email{rmendez@u.uchile.cl}
\author{Elliott P. Horch\footnote{Adjunct Astronomer, Lowell Observatory} }
\affil{Department of Physics, Southern Connecticut State University, 501 Crescent Street, New Haven, CT 06515, USA}
\email{horche2@southernct.edu}

\begin{abstract}
The  results  of  speckle  interferometric observations  at  the  SOAR
telescope  in  2015  are  given,  totaling 1303  measurements  of  924
resolved binary and multiple stars and non-resolutions of 260 targets.
The separations  range from  12\,mas to 3\farcs37  (median 0\farcs17);
the maximum measured magnitude difference  is 6.7 mag.  We resolved 27
pairs for the first time, including 10 as inner or outer subsystems in
previously known  binaries, e.g.   the 50-mas pair  in $\epsilon$~Cha.
Newly resolved pairs are  commented upon.  We discuss three apparently
non-hierarchical systems discovered in this series, arguing that their
unusual configuration results  from projection. The resolved quadruple
system HIP~71510 is studied as well.
\end{abstract} 

\section{Introduction}
\label{sec:intro}

We report  here a  large set of  binary-star measurements made  at the
4.1-m  Southern  Astrophysical  Research  Telescope  (SOAR)  with  the
speckle camera,  HRCam.  This paper continues the  series published by
\citet[][hereafter   TMH10]{TMH10},  \citet{SAM09},  \citet{Hrt2012a},
\citet{Tok2012a},      \citet{TMH14},      and      \citet[][hereafter
  SOAR14]{TMH15}.   Our  primary  goals  are characterization  of  the
binary  and  multiple  star  population  in  the  solar  neighborhood,
improvement  of  known  orbital  elements, and  determination  of  new
orbits.    Taking   advantage   of   the   high   angular   resolution
($\sim$30\,mas), we  follow close pairs  with fast orbital  motion and
periods of  a few  years or decades;  measurements of other  wider and
slower  binaries are  useful for  future orbit  determination  and for
checking internal data  consistency.  A subset of the  targets we have
observed are drawn from the {\it Hipparcos} \citep{HIP} and Geneva-Copenhagen
\citep{N04}  catalogs.  As  such,  these   represent  a  Southern
Hemisphere complement to recent work at Lowell Observatory's Discovery
Channel Telescope \citep{Horch2015}.

Orbits of binary stars serve a variety of astrophysical programs. Data
on masses  are still needed for stars  of high and low  masses, over a
range  of metallicities and  ages (e.g.   pre-main-sequence). Accurate
parallaxes from {\it Gaia}  combined with accurate orbits derived from
long-term  ground-based monitoring  will greatly  advance  the current
census  of stellar  masses.  Orbits  of hierarchical  multiple systems
give insights on their origin and evolution; the same is true for multiple
systems  containing  exo-planets and  debris  disks.  We also  provide
differential photometry of close pairs.

\section{Observations}
\label{sec:obs}

\subsection{Instrument and observing method}

The   observations  reported   here  were   obtained  with   the  {\it
  high-resolution camera} (HRCam) -- a fast imager designed to work at
the  4.1-m SOAR  telescope  \citep{TC08}. For  practical reasons,  the
camera was  mounted on  the SOAR Adaptive  Module \citep[SAM,][]{SAM}.
However,  the laser  guide star  of SAM  was not  used (except  in May
2015); the  deformable mirror of  SAM was passively flattened  and the
images are  seeing-limited.  The  SAM module corrects  for atmospheric
dispersion and helps  to calibrate the pixel scale  and orientation of
HRCam (see SOAR14).  The  transmission curves  of
HRCam   filters  are   given  in   the   instrument  manual.\footnote{
  \url{http://www.ctio.noao.edu/soar/sites/default/files/SAM/\-archive/hrcaminst.pdf}}
We  used  mostly  the  Str\"omgren  $y$ filter  (543/22\,nm)  and  the
near-infrared $I$ filter (788/132\,nm).

\subsection{Observing runs}

The observing time for this  program was allocated through NOAO (three
nights,  programs 15A-0097  and 15B-0009,  PI A.T.)  and by  the Chilean
National  Time Allocation  Committee  (three nights  in 2015B,  program
CN2015B-6, PI  R.A.M.).  All  observations were made  by A.T.,  sharing the
allocated time  between both  programs to cover  the whole sky  and to
improve  temporal cadence for pairs with fast motion.

\begin{deluxetable}{ l l r r c } 
\tabletypesize{\scriptsize}    
\tablecaption{Observing runs
\label{tab:runs} }                    
\tablewidth{0pt}     
\tablehead{ \colhead{Run}  &
\colhead{Dates}  & 
\colhead{$\theta_0$} & 
\colhead{Pixel} &
\colhead{$N_{\rm obj}$} \\
 &   &  \colhead{(deg)} & \colhead{ (mas) } & 
}
\startdata
1 & 2015 Feb 7       & 0.30     & 15.23 & 64  \\ 
2 & 2015 Mar 2-3     & $-0.30$  & 15.23 &  207  \\ 
3 & 2015 Apr 1       & $-0.60$  & 15.23 & 44  \\ 
4 & 2015 May 2-3     &  $-2.74$ & 15.23 & 195  \\ 
5 & 2015 Jun 30      &  $-2.60$ &  15.23 & 126  \\ 
6 & 2015 Jul 16     & $-2.60$   & 15.23 &  126 \\   
7 & 2015 Sep 26,28  &  $-2.68$  & 15.23 &  323 \\ 
8 & 2015 Nov 27-29  &  $-2.84$  & 15.19 &  ~425 
\enddata
\end{deluxetable}

Table~\ref{tab:runs}  lists   the  observing  runs,   the  calibration
parameters (position angle offset  $\theta_0$ and pixel scale in mas),
and  the number of  objects covered  in each  run. The  calibration of
angle  and scale  was  done with  respect  to wide  pairs with  linear
motion,  as explained  in SOAR14.  It  was revisited  and improved  by
including more calibrators and the  2015 data. The scale is consistent
within 0.2 per cent, the  calibration of position angle is accurate to
0\fdg1.

In  2015A,  the NOAO  TAC  allocated  only  two half-nights  for  this
program. Additional  observing time was  granted by the  SOAR director
during engineering nights  or half-nights. The two nights  in May (Run
4)  were  dedicated  to  the  follow-up  of  {\it  Kepler-2}  exo-host
candidates  by  the Yale  University  team;  measurements of  binaries
observed when  the main targets  were out of visibility  are published
here, while  the main results are presented  by \citet{Schmitt2016}. In
this Run,  the SAM  AO system with  UV laser  was used to  improve the
image  quality  and, hence,  magnitude  limit  on  faint {\it  Kepler}
targets.  The faintest binaries were  observed with the laser as well.
The gain in sensitivity from using the laser depends on the turbulence
profile  and other  factors,  so  it is  difficult  to quantify.   The
three-night Run 7  was allocated for the program of  R.A.M.  Two
nights  of  this Run  were  lost to  bad  weather,  but an  additional
engineering half-night  was granted to  finish the program.   The last
two-night Run  8 in 2015 November  enjoyed clear sky  and good seeing.
It was preceded by a half-night of engineering observations taken with
the malfunctioning SAM instrument, with strongly aberrated images.

\begin{figure}[ht]
\epsscale{1.0}
\plotone{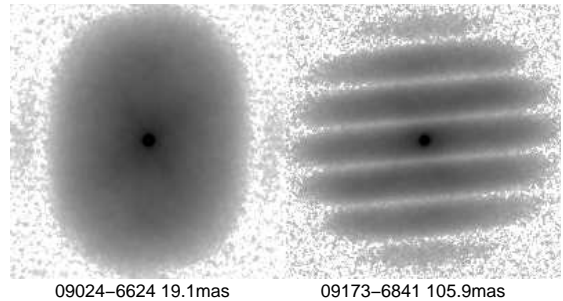}
\caption{\label{fig:ref} Use of a binary  star as a reference. The power
  spectra in logarithmic scale  are displayed for two binaries located
  in the  same part of the  sky ($y$ band, 2-ms  exposures). The close
  binary  on  the  left  with  a  separation  of  19\,mas  (below  the
  diffraction  limit) is  fitted better  if  the binary  on the  right
  (106\,mas) is used as a reference.  }
\end{figure}

\begin{figure}[ht]
\epsscale{1.0}
\plotone{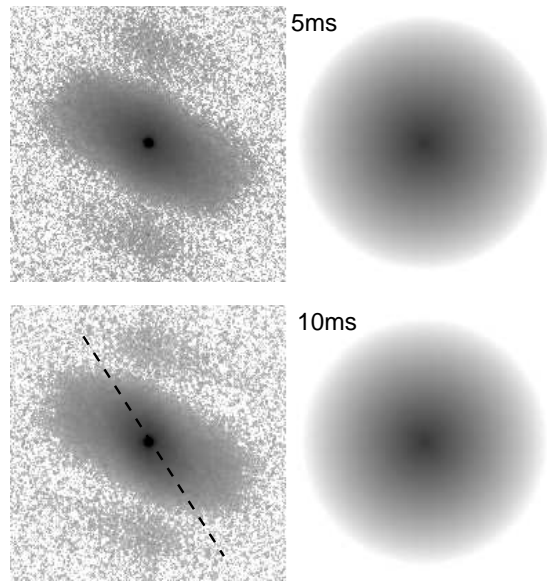}
\caption{\label{fig:vib}  Influence  of  telescope vibrations  on  the
  results.   The power  spectra of  TOK~387  (04386$-$0920, separation
  40\,mas)  taken in  2015.9 with  5-ms  and 10-ms  exposure time  are
  displayed  on logarithmic  scale,  with corresponding  ``synthetic''
  references on the right side.  The 10-ms data are partially affected
  by  telescope  vibration that  elongated  the  central  zone of  the
  spectrum  in the  direction of  the dashed  line. As  a  result, the
  fitted position  angle is  biased by $6^\circ$  for the  10-ms data,
  which were discarded.  Use of real reference spectra  taken with the
  same exposure  helps to  account for vibration-induced  asymmetry of
  the power spectra, avoiding such biases.  }
\end{figure}

\subsection{Data processing}

The procedure for processing data is described in TMH10. As a first step, power
spectra and average re-centered images are calculated from the data
cubes. The auto-correlation functions (ACFs) are computed from the
power spectra. They are used to detect companions and to evaluate the
detection limits. The parameters of binary and triple stars are
determined by fitting the power spectrum to its model, which is a
product of the binary (or triple) star spectrum and the reference
spectrum. We used as a reference the azimuthally-averaged spectrum of
the target itself in the case of binaries wider than 0\farcs1. For
closer pairs, the ``synthetic'' reference was used (see TMH10). 

We added  a third option of  using observations of  single or resolved
binary stars as a reference. In  the latter case, the signature of the
binary is  removed from  the reference spectrum  with the help  of the
previously  fitted binary  parameters.  This  works for  binaries with
$\Delta m \ge 1$ mag,  otherwise the binary fringes have high contrast
and  their  removal  by  division  increases the  noise.  The  use  of
reference stars  helps in fitting  difficult cases such as  very faint
and/or  close companions.   Figure~\ref{fig:ref}  shows the  elongated
power spectrum  of the  very close pair  TOK~197 with a  separation of
19.1\,mas (under the diffraction limit) observed in 2015.9.  The wider
0\farcs106  pair FIN~363  in the  same area  of the  sky  was observed
shortly  after,  with the  same  $y$ filter  and  with  the same  2-ms
exposure.   Both data  cubes  are of  excellent  quality, showing  the
speckle  signal out  to the  cutoff  frequency. The  use of  reference
stars improves the  quality of  the fit (a  smaller $\chi^2/N$)  and assures
that  any remaining asymmetric  distortions of  the spectrum,  e.g. by
telescope  vibration,  are  properly  modeled  (Figure~\ref{fig:vib}).
Observing a  large number of binaries with  standardized exposure time
and filters helped to find references for many targets.
 We  used  reference stars when  they  were available  and when  they
reduced  the   $\chi^2/N$  error  metric  compared   to  the  standard
self-reference technique  described in TMH10.   The difference between
measures of the  same object in the $y$ and  $I$ filters was generally
reduced with reference stars.

Wide binaries  resolvable in the re-centered  long-exposure images are
processed  by another  code that  fits only  the  magnitude difference
$\Delta m$,  using the binary  position from speckle  processing. This
procedure  corrects   the  bias  on  $\Delta  m$   caused  by  speckle
anisoplanatism  and establishes the  correct quadrant  (flag *  in the
data table).

We also  calculated shift-and-add (SAA or  ``lucky'') images, centered
on the  brightest pixel in  each frame and weighted  proportionally to
the  intensity  of that  pixel.   All  frames  without rejection  were
co-added.   Binary companions,  except  the closest  and the  faintest
ones,  are detectable  in these  SAA images,  helping to  identify the
correct quadrant.  Such cases  are marked  by the flag  q in  the data
table.  Quadrants of  the remaining binary stars are  guessed based on
prior data or orbits, not measured directly.

\section{Results}
\label{sec:res}

\begin{figure*}[ht]
\epsscale{1.1}
\plotone{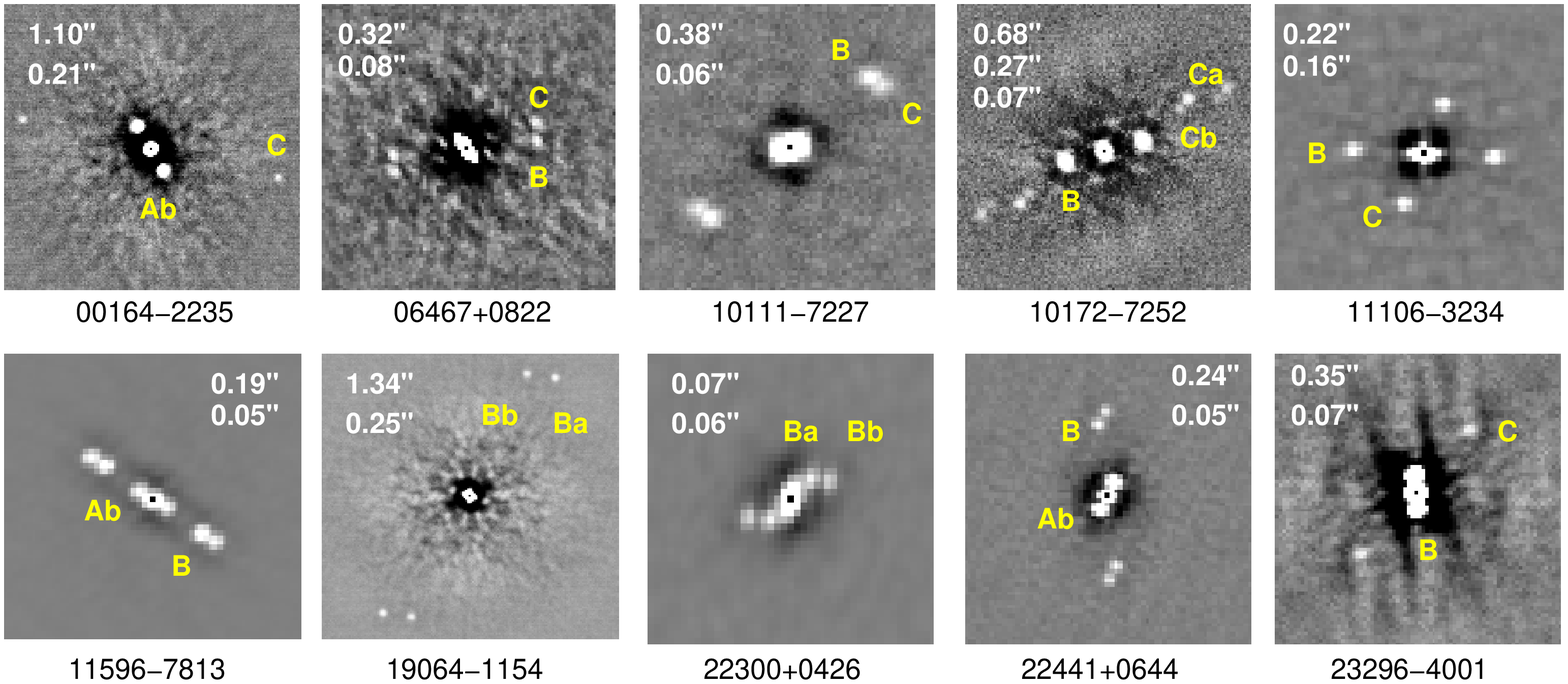}
\caption{\label{fig:ACF}  Fragments of ACFs  of newly  resolved triple
  systems (North  up, East  left, arbitrary scale).  Peaks  in the
  ACFs are labeled by component designations. Angular separations in
  the wide and close pairs are listed in the images. }
\end{figure*}

\subsection{Data tables}

\begin{deluxetable*}{l l l  ccc  rc cc l r r l }                                                                                                                                
\tabletypesize{\tiny}   
\tablenum{2}                                                                                                                                                  
\tablecaption{Measurements of double stars at SOAR  (fragment)                                                                                                                              
\label{tab:double} }                                                                                                                                                            
\tablewidth{0pt}                                                                                                                                                                
\tablehead{                                                                                                                                                                     
\colhead{WDS} & \colhead{Discoverer} & \colhead{Other} & \colhead{Epoch} & \colhead{Filt} & \colhead{N} & \colhead{$\theta$} & \colhead{$\rho \sigma_{\theta}$} &               
\colhead{$\rho$} & \colhead{$\sigma \rho$} & \colhead{$\Delta m$} & \colhead{[O$-$C]$_{\theta}$} & \colhead{[O$-$C]$_{\rho}$} & \colhead{Reference} \\         
\colhead{(2000)} & \colhead{Designation} & \colhead{name} & +2000 & & & \colhead{(deg)} & (mas) & ($''$) & (mas) & (mag) & \colhead{(deg)} & \colhead{($''$)}                   
& \colhead{code\tablenotemark{a}}  }   
\startdata     
00006$-$5306 &	HJ 5437 &	HIP 50 &	15.7434 &	I &	2 &	337.3 &	0.3 &	1.4110 &	1.1 &	2.7 *  &  &   &   \\ 
00008+1659 &	BAG 18 &	HIP 68 &	15.9127 &	I &	2 &	343.9 &	0.4 &	0.7063 &	0.4 &	2.7 *  &  &   &   \\ 
00024+1047 &	A 1249 AB &	HIP 190 &	15.9127 &	I &	2 &	247.2 &	0.1 &	0.2808 &	0.2 &	0.8 q &	179.3 &	0.046 &	Zir2003  \\ 
00036$-$3106 &	TOK 686 &	HIP 290 &	15.7382 &	I &	3 &	5.3 &	1.0 &	0.1257 &	1.9 &	3.2    &  &   &   \\ 
            &	       &	           &	15.9130 &	I &	2 &	6.5 &	2.2 &	0.1160 &	17.3 &	3.1 :  &  &   &   \\ 
00058$-$6833 &	HDS 4 &	HIP 488 &	15.9100 &	I &	2 &	226.0 &	1.0 &	0.1920 &	0.3 &	1.7 q  &  &   &   \\ 
00061+0943 &	HDS 7 &	HIP 510 &	15.7433 &	I &	2 &	12.3 &	0.3 &	0.2515 &	0.4 &	0.2    &  &   &   \\ 
00071$-$1551 &	HDS 11 &	HIP 584 &	15.7382 &	I &	2 &	261.9 &	1.6 &	0.4093 &	0.7 &	2.6 q  &  &   &   \\ 
00090$-$5400 &	HDO 181 &	HIP 730 &	15.7434 &	I &	2 &	20.9 &	0.2 &	0.3655 &	0.6 &	2.1 q &	$-$4.5 &	$-$0.014 &	Ary2002b  
\enddata 
\tablenotetext{a}{References to VB6 are provided at  \url{http://ad.usno.navy.mil/wds/orb6/wdsref.txt} }
\end{deluxetable*}

\begin{deluxetable*}{l l l   c c c  c c c c  }                                                     
\tabletypesize{\scriptsize}               
\tablenum{3}                                                      
\tablecaption{Unresolved stars  (fragment)                                                                 
\label{tab:single} }                                                                            
\tablewidth{0pt}                                                                                
\tablehead{                                                                                     
WDS (2000) & \colhead{Discoverer} & \colhead{Hipparcos} & \colhead{Epoch} & \colhead{Filter} &  
\colhead{N} & \colhead{$\rho_{\rm min}$} &    \multicolumn{2}{c}{5$\sigma$ Detection Limit} & $\Delta m$   \\             
$\alpha$, $\delta$ (J2000)  & \colhead{Designation} & \colhead{or other} & \colhead{+2000} & & &                           
& \colhead{$\Delta m (0\farcs15)$}  & \colhead{$\Delta m (1'')$}    & flag    \\             
 & \colhead{or other name}  & \colhead{name}  & & & & (arcsec) & \colhead{(mag)} &  \colhead{(mag)}  &     
}                                                                                               
\startdata     
00012$-$0005 &	TOK 359 &	HIP 93 &	15.7433 &	I &	3 &	0.040 &	3.95 &	5.07  \\ 
00036$-$3106 &	TOK 686 &	HIP 290 &	15.9130 &	y &	2 &	0.028 &	4.87 &	6.41  \\ 
00039$-$2824 &	HIP 305 &	HIP 305 &	15.7382 &	I &	2 &	0.040 &	4.35 &	5.26  \\ 
00052$-$6251 &	HIP 425 &	HIP 425 &	15.7434 &	I &	2 &	0.040 &	4.14 &	4.99  \\ 
00138+0812 &	HIP 1103 &	HIP 1103 &	15.7433 &	I &	2 &	0.040 &	4.03 &	5.34  \\ 
00194$-$4007 &	HIP 1552 &	HIP 1552 &	15.9130 &	I &	2 &	0.040 &	3.97 &	6.37  \\ 
00219$-$2300 &	HJ 1957 A &	HIP 1732 &	15.7382 &	I &	2 &	0.040 &	4.17 &	5.64  \\ 
00221$-$2643 &	HIP 1746 &	HIP 1746 &	15.9130 &	I &	2 &	0.040 &	4.04 &	5.51  
\enddata 
\end{deluxetable*}

The data tables have almost the  same format as in the previous papers
of  this  series. They  are  available  in  full only  electronically.
Table~2  lists  1303  measures   of  924  resolved  binary  stars  and
subsystems, including 27 newly  resolved pairs. The columns of Table~2
contain  (1) the  WDS  \citep{WDS} designation,  (2) the  ``discoverer
designation'' as adopted  in the WDS, (3) an  alternative name, mostly
from the {\it Hipparcos}  catalog, (4) Besselian epoch of observation,
(5)  filter,  (6) number  of  averaged  individual  data cubes,  (7,8)
position angle  $\theta$ in degrees and internal  measurement error in
tangential direction $\rho  \sigma_{\theta}$ in mas, (9,10) separation
$\rho$ in  arcseconds and its  internal error $\sigma_{\rho}$  in mas,
and  (11) magnitude  difference  $\Delta m$.  An  asterisk follows  if
$\Delta  m$ and  the true  quadrant are  determined from  the resolved
long-exposure image;  a colon  indicates that the  data are  noisy and
$\Delta m$ is likely over-estimated  (see TMH10 for details); the flag
``q'' means the quadrant is  determined from the SAA image.  Note that
in the cases of multiple  stars, the positions and photometry refer to
the  pairings  between  individual  stars, not  the  photo-centers  of
subsystems.

For  stars with known  orbital elements,  columns (12--14)  of Table~2
list the residuals to the  ephemeris position and code of reference to
the  orbit adopted in  the Sixth  Catalog of  Orbits of  Visual Binary
Stars            \citep[][hereafter            VB6]{VB6}.\footnote{See
  \url{http://ad.usno.navy.mil/wds/orb6/wdsref.html}}

Table~3 contains the  data on 260 unresolved stars,  some of which are
listed   as  binaries   in  the   WDS  or   resolved  here   in  other
filters. Columns (1) through (6)  are the same as in Table~2, although
Column (2)  also includes other  names for objects  without discoverer
designations.   For  stars  that  do  not have  entries  in  the  WDS,
fictitious WDS-style codes  based on the J2000 position  are listed in
Column (1).  Column (8) is the estimated resolution limit equal to the
diffraction  radius $\lambda/D$.   Columns (8,9)  give the  $5 \sigma$
detection  limits $\Delta  m_5$ at  $0\farcs15$ and  $1''$ separations
determined by the procedure described in TMH10.  When two or more data
cubes are processed, the largest  $\Delta m$ value is listed. The last
column marks by colons noisy  data mostly associated with faint stars.
In such cases, the quoted  $\Delta m$ might be too large (optimistic);
however,  the  information  that  these  stars were  observed  and  no
companions were found is still useful for statistics.

\subsection{Newly resolved pairs}
\label{sec:new}

\begin{deluxetable*}{l l l  cc  r r l lr }                                                                                                                                
\tabletypesize{\scriptsize}                                                                                                                                                     
\tablenum{4}
\tablecaption{Newly resolved   pairs
\label{tab:new} }                    
\tablewidth{0pt}  
\tablehead{
\colhead{WDS} & \colhead{Discoverer} & \colhead{Other} & \colhead{Epoch}  &  \colhead{Filt} &\colhead{$\theta$} &                
\colhead{$\rho$}  & \colhead{$\Delta m$} & \colhead{Sp.} &  \colhead{$\pi_{\rm HIP2}$}  \\         
\colhead{(2000)} & \colhead{Designation} & \colhead{name} & +2000 & &
\colhead{(deg)} &  ($''$) &  (mag) & type & (mas) }                       
\startdata 
00036$-$3106 &	TOK 686 &	HIP 290 &	15.7382 &	I &	5.3 &	0.1257 & 3.2  & F9V  & 15.4 \\
00164$-$2235 &	HDS 36 Aa,C &	HIP 1306 &	15.7382 &	I &	254.5 &	1.1000 & 4.9  & G3V  & 13.1 \\
01077$-$1919 &	TOK 687 &	HIP 5291 &	15.7382 &	I &	41.8 &	0.6077 & 4.8  & G8V  & 11.4 \\
01380+0946 &	TOK 688 &	HIP 7604 &	15.7383 &	I &	13.4 &	0.0559 & 2.2  & F5   & 5.1 \\
03322$-$3134 &	TOK 691 &	HIP 16481 &	15.7435 &	I &	312.6 &	0.1656 & 3.9  & F0V  & 3.9 \\
06467+0822 &	HDS 940 BC &	HIP 32475 &	15.9081 &	y &	2.4 &	0.0828 & 0.3  & F0IV & 14.6 \\ 
10111$-$7227 &	HDS 1468 BC &	HIP 49879 &	15.1688 &	I &241.2   &	0.0631 & 0.3  & F3V   & 7.0 \\ 
10172$-$7252 &	HEI 494 A,Ca &	HIP 50381 &	15.2502 &	I &302.1 &	0.6793 & 2.9   & A8V   & 5.2 \\
10172$-$7252 &	HEI 494 Ca,Cb &	HIP 50381 &	15.2502 &	I &153.1 &	0.1020 & 0.8  &  A8V   & 5.2 \\
10269$-$5340 &	TOK 693 Aa,Ab &	HIP 51144 &	15.1714 &	I &322.6 &	0.3464 & 2.1   & F6V    & 4.8 \\
11106$-$3234 &	I 213 AC &	HIP 54611 &	15.1686 &	I &156.9 &	0.1687 & 3.1  &  F5IV & 6.8 \\
11596$-$7813 &	HJ 4486 Aa,Ab &	$\epsilon$ Cha&	15.2502 &	y &57.3 &	0.0512 & 0.2  &  B9V    & 9.0 \\
16004$-$5107 &	TOK 694 &	HD 143055 &	15.1689 &	I &219.5 &	0.0246 & 1.5 : & F0V    & \ldots \\
18040+0150 &	TOK 695 &	HIP 88481 &	15.3358 &	y &26.5 &	0.0608 & 1.8 q & G5    & 29.0 \\
19064$-$1154 &	RST 4028 Ba,Bb &HIP 93827 &	15.7374 &	I &78.2 &	0.2520 & 0.0  & M0V    & 26.3  \\
19512$-$7248 &	TOK 697 Aa,Ab &	HIP 97690 &	15.3359 &	y &30.9 &	0.0212 & 1.0 : & F6V    & 11.5 \\
19563$-$3137 &	TOK 698 &	HIP 98108 &	15.5410 &	I &58.9 &	0.0801 & 1.1 q & G3V    & 17.0 \\
20048+0109 &	TOK 699 &	HIP 98878 &	15.4973 &	I &317.7 &	0.1571 & 2.5 q &  G5    & 26.1 \\
20286$-$0426 &	TOK 700 &	HIP 100998 &	15.4972 &	I &222.4 &	0.1081 & 0.6 q &  F8    & 15.3\\ 
20574$-$5905 &	TOK 701 Ba,Bb&	HIP 103438B&	15.5411 &	I &258.9 &	0.1496 & 0.7 q  & G5V?   & 18.6  \\ 
22139$-$2216 &	TOK 702 &	HIP 109753 &	15.7379 &	I &39.2 &	1.7792 &4.9 *  & G5V  & 12.0 \\
22300+0426 &	STF 2912 Ba,Bb &HIP 111062 &	15.9127 &	y &268.4 &	0.0599 & 0.8  & F8+A4  & 19.3 \\
22308$-$2410 &	HDS 3192 Aa,Ab &HIP 111133 &	15.5413 &	I &149.8 &	0.0545 & 0.4 q & K1III  & 3.5 \\
22441+0644 &	TOK 703  &	HIP 112240 &	15.4974 &	I &54.2 &	0.0380 & 0.8   &  F5    & 17.1 \\ 
23224$-$4636 &	CPO 637 Aa,Ab &	HIP 115386 &	15.9128 &	I &48.8 &	1.1456 &5.2 *  &  G5V & 15.3    \\ 
23228+2034 &	TOK 704 Ba,Bb &HIP 115417B &	15.7379 &	I &181.2 &	0.7470 & 2.0 q  & K0V? & 26.8      \\ 
23296$-$4001 &	HDS 3346 AC &	HIP 115957 &	15.9128 &	I &315.9 &	0.3494 & 4.3    & G6V & 6.8 
\enddata
\end{deluxetable*}

Table~4  lists 27  newly  resolved  pairs.  Its  format  is similar  to
that of Table~2. For  some multiple systems,  we used existing  discover codes
and simply added new component  designations.  The last two columns of
Table~4 contain  the spectral  type (as given  in SIMBAD  or estimated
from   absolute   magnitude)   and   the  {\it   Hipparcos}   parallax
\citep[][hereafter HIP2]{HIP2}.   Fragments of ACFs  of newly resolved
triple systems are shown  in Figure~\ref{fig:ACF}. We comment on these
objects  below. The  following abbreviations  are used:  PM  --- proper
motion, CPM  --- common proper motion,  RV --- radial  velocity, SB1 and
SB2  --- single- and  double-lined spectroscopic  binaries.
Orbital  periods are  estimated from  projected separation  as  $P^* =
(\rho /p)^{3/2} M^{-1/2}$, where $\rho$ is the angular separation, $p$
is  parallax,  $M$  is the  mass  sum,  and  $P^*$  is the  period  in
years.  Data  from  the  spectroscopic Geneva-Copenhagen  Survey,  GCS
\citep{N04} are used for some targets.

{\it 00036$-$3106.}  HIP 290  is a binary  detected by  its astrometric
acceleration \citep{MK05} and variable  RV (GCS). The separation implies an
orbital period of $P^* = 17$\,yr.

{\it  00164$-$2235.}   The faint  tertiary  companion  to HIP~1306  at
1\farcs1   was   first    resolved   on   direct   and   coronagraphic
adaptive-optics (AO) images by  \citet{Boccaletti2004}, although this is not
reflected in the  WDS.  They quote a separation  of 1\farcs075 and a
magnitude difference of  3.5 mag at 2.12\,$\mu$m between  A and C, but
do  not list the  position angle.   Looking at  their Figure~9,  it is
similar to 254$^\circ$  measured here, while we find  $\Delta I = 4.9$
mag.

{\it  01077$-$1919.}  HIP~5291  (HD  6720) was  observed  because the GCS
detected  double lines  (SB2).  However,  it is  likely  that the  new
0\farcs6 companion with $\Delta I = 4.8$ mag is a tertiary.

{\it 01380+0946.}  HIP~7604  (HD~10016) is listed in the  GCS as an SB
with a mass ratio of 0.960$\pm$0.010 and [Fe/H] of $-$0.20. Apparently
the SB is resolved here and  shows a rapid motion.  It was observed by
lunar  occultations   but  not  resolved   \citep{Richichi2006},  also
unresolved   by   speckle  interferometry   at   the  WIYN   telescope
\citep{Horch2011}.

{\it 03322$-$3134.}  HIP~16481 (HD~22054) is  an SB2 with a  period of
10.1  days, equal-mass components,  and a  small eccentricity  of 0.13
\citep{N97}.  Obviously, the 0\farcs17 interferometric companion found
here  cannot correspond  to the  spectroscopic one,  so the  system is
triple.

{\it  06467+0822.}   HIP~32475  (HD  49015)  is  a  variable  star  of
$\gamma$~Dor  type V839~Mon, an  X-ray source,  and a  0\farcs3 visual
binary HDS~940.  It was observed by speckle interferometry at the WIYN
telescope several  times \citep[e.g.][]{Horch2008}.  Re-examination of
the WIYN data  by one of us (E.H.) shows doubling  of the secondary in
2005, 2007, and 2012,  confirming its resolution here.  The separation
of  BC  implies  an  orbital  period of  $P^*  \sim  10$\,yr;  further
observation and archival  data will soon lead to  the determination of
its orbit.

{\it  10111$-$7227.} The  secondary component  of HDS~1468  is resolved
into the close 63-mas  pair BC.  Its  projected  separation  implies $P^*  \sim
20$\,yr. Indeed, the pair BC is in rapid retrograde motion, turning by
$16^\circ$ during 2015.

{\it 10172$-$7252.} A new component  C to HEI~494~AB is discovered at a
separation  of 0\farcs67. This  component itself  is also  a 0\farcs07
binary Ca,Cb.  The  resolution is confirmed by re-observation  in the following
runs.   The relative  position  and magnitude  difference  of Ca,Cb are
measured only crudely.  If C is  physical to AB (the object has a low
galactic  latitude, so  the  sky is  crowded),  this is  a new  2+2
quadruple. 

{\it 10269$-$5340.}  HIP  51144 is an SB2 according to  the GCS. It was
resolved for  the first time  at 0\farcs37 in  Run 2 and  confirmed in
Runs 3 and 9. The projected separation implies $P^* = 430$\,yr, so the
new  companion is  not the  one causing  double lines.  The  system is
likely  triple.  Using  {\it Hipparcos}  photometry,  \citet{Koen2002}
found periodic variability with $P=1.33$ days and an amplitude of 0.03
mag.

{\it  11106$-$3234.} The  pair I~213,  known as  a binary  since 1879,
unexpectedly turned out to be  a triple system in a nearly equilateral
triangular  configuration, with  two  components B  and  C of  similar
brightness (both about 3 mag fainter than A).  It was confirmed on the
following night, showing that this is not a transient.  This system is
discussed in the next sub-section.

{\it  11596$-$7813.}  This  is  the  B9V  star  $\varepsilon$~Chameleontis
(HR~4583), the  most massive member  of the association bearing  its name.
Known as a binary HJ~4486~AB since 1835, it turns out to be a triple
system composed  of three  similar stars.  It  is surprising  that the
inner 50-mas  pair Aa,Ab was not discovered  earlier in high-resolution
surveys  of the  $\varepsilon$~Cha association.   The  separation implies
$P^* \sim 7$\,yr, so  rapid motion of Aa,Ab is expected.  The fringes
of Aa,Ab  have unit  contrast.  Processed as  a binary,  the subsystem
Aa,Ab has  magnitude difference $\Delta y  = \Delta I =  0.2 \pm 0.05$
mag, while the triple-star fit gives a slightly larger $\Delta m$. The
components  Aa and B  are also  equal and  our photometry  agrees with
$\Delta Hp_{\rm AB}  = 0.69$ mag measured by  {\it Hipparcos}. Note the
linear  configuration of this  triple in  Figure~\ref{fig:ACF}. The
outer pair AB was at $(179^\circ, 1\farcs6)$ in 1835 and has now closed
down to $(235^\circ,  0\farcs19)$. Its nearly radial motion suggests a
highly inclined or eccentric orbit.  Study  of this interesting triple system can
give some  clues to understanding the peculiarity  of this association
which lacks both low-mass stars and wide binaries according to \citet{Becker2013}.

{\it  16004$-$5107.}  HD  143055 (F0V)  was observed  on  request from
K.~Helminiak.   The  elongated   power  spectrum  indicates  tentative
resolution,  confirmed  in  another  run. However,  the  measures  are
of low accuracy.

{\it  18040+0150.}  According  to  D.~Latham  (private  communication,
2012), HIP~88481 = HD~165045 is an  SB2 with a period of 1.6\,yr and a
mass ratio of 0.6; its low mass function suggests a face-on orbit. The
expected semi-major  axis is 46\,mas,  in agreement with  the measured
separations  and  the fast  motion  of 110$^\circ$~yr$^{-1}$  observed
over three runs.

{\it  19064$-$1154.} The  faint secondary  of HIP~93827  (HD~177758) is
resolved into  a 0\farcs25 pair  of equal stars. Their  estimated masses
are 0.5 ${\cal M}_\odot$,  the period of Ba,Bb is $P^* \sim 30$\,yr. This is a
solar-type triple system within 67\,pc. 

{\it  19512$-$7248.} HIP~97690 =  HD~186502 (F6V)  is an SB2  according to
the GCS. \citet{Gln2007}  computed astrometric orbit  with $ P=2.877$\,yr,
implying a semi-major axis of 28\,mas. While the separations measured in
three runs  match the expected  axis, the angles  disagree  with the
astrometric orbit; the pair moves very fast.

{\it 19563$-$3137.} HIP~98108 = HD~188432 (G3V) has double lines (GCS) and
astrometric acceleration. The 80-mas separation implies a short
period $P^* \sim 7$\,yr. The star was not resolved in a survey of
astrometric binaries at Gemini-South \citep{Tok2012b}.

{\it   20048+0109.}   HIP~98878  = HD~190412 is a spectroscopic triple
with periods of 251\,days and 7.8\,yr (D.~Latham, private
communication, 2012). The outer period corresponds to the axis of
0\farcs10, and the tertiary was resolved at 0\farcs16. Astrometric
acceleration was also detected. The star is on the exoplanet search
program at the Keck telescope.

{\it  20286$-$0426.} The GCS suspected that  HIP~100988 has a variable
RV. Its resolution at 0\farcs10 implies an orbital period $P^* \sim
10$\,yr. 

{\it 20574$-$5905.}  The 4\farcs6 pair  COO~241~AB (HIP~103438) belongs
to  the 67-pc  sample.   We observed  both  components separately  and
resolved the  secondary into a  new 0\farcs15 pair Ba,Bb  with estimated
period $P^* \sim  20$\,yr. The outer pair has a  much longer period of
$\sim$40\,kyr.   \citet{Desidera2006}  detected  an RV  difference  of
2\,km~s$^{-1}$  between A and  B, which  can partly  be caused  by the
motion of Ba,Bb.  The pair Ba,Bb  is more massive than the single star
A; it  is located above  the main sequence  as expected for  a binary.

{\it 22139$-$2216.} HIP~109753 is an SB2 in the GCS. The faint companion at
1\farcs8 found here could be a tertiary if it is not optical. 

{\it 22300+0426.}  This bright star  37~Peg (HR 8566) is  a well-known
visual binary  STF~2912 with an  orbital period of 125  years. Despite
extensive literature, including speckle  coverage, the binarity of the
secondary component has never been suspected. Here it is resolved into a
close  pair  which, projected  on  the  primary,  forms an  apparently
non-hierarchical  configuration. The  system is  further  discussed in
Section~\ref{sec:trapezia}.

{\it 22308$-$2410.} The bright  star HIP~111133 (K1III) was resolved by
{\it Hipparcos} at 0\farcs2 and never confirmed since. Its observation
at  SOAR shows  a  change in  angle  by $6^\circ$,  while the  primary
component is a new 53-mas pair Aa,Ab. The fringes of Aa,Ab are of high
contrast, processing it  as a binary gives $\Delta y =  \Delta I = 0.2
\pm  0.05$ mag,  while the  triple-star  fit gives  a slightly  larger
$\Delta   m  \sim   0.4$ mag.   The   period  of   Aa,Ab   is  $P^*   \sim
40$\,yr. Considering their  similar colors, both Aa and  Ab are giants
at a similar evolutionary stage.

{\it 22441+0644.} The acceleration  binary HIP~112240 (F5) is resolved
at 38\,mas, implying a period of $P^* \sim 2$\,yr.  The RV variability
was detected by the GCS.  It was not resolved in the following run and
tentatively  resolved again in  2015.7 and  2015.9.  All  measures are
derived  by fitting the  elongated  power spectrum  (no  second fringe  is
detected), so they are of questionable accuracy. 

{\it 23224$-$4636.} HIP~115386 belongs to the 67-pc sample. It has a
CPM companion at 38\farcs5. Both components were observed at
SOAR. While B was unresolved, we found a new faint pair AC at
1\farcs14. The sky is not crowded, so AC is likely physical.

{\it  23228+2034.}   HIP~115417B  was  resolved  at  the  Gemini-North
telescope  in  2015  (E.H.  \& A.T., in  preparation).   It  is
confirmed here,  while the  A-component is unresolved.   The subsystem
Ba,Bb  causes a  ``wobble''  with  a period  of  $\sim$115\,yr in  the
observed motion of the 5\farcs9 pair AB.

{\it 23296$-$4001.} The known  close binary HIP~115987 was observed in
the $y$ and $I$ bands. The  new faint component C at 0\farcs35 is seen
only in $I$.

\subsection{Comments on other pairs}
\label{sec:other}

{\it 02418$-$5300.}  The close pair  CD$-$53\degr~544 was first  resolved by
\citet{Elliott2015} in a survey  of young stars (see also 02305$-$4342
= ELP~1 and 08138$-$0738 =  ELP~2 measured here). However, no measures
of this pair were published in the discovery paper, so the designation
TOK~690~Aa,Ab is adopted here. 

{\it 04074$-$4255.} The primary of HDS 522 has spectral type K0III: 
the secondary is bluer, which explains its non-resolution in the $I$ filter.

{\it 13401$-$6033.} New measurement confirms that the 0\farcs9 pair BC
is physical, despite crowdedness of the sky in this area.   

\subsection{Apparently non-hierarchical triples}
\label{sec:trapezia}

The  discovery  of the  triple  system  HIP~54611 (WDS  J11106$-$3234,
I~213)    in   a    nearly   equilateral    triangular   configuration
(Figure~\ref{fig:ACF}) raises the question of its dynamical stability.
Small $N$-body systems with  comparable separations decay during a few
crossing   times,   leaving   binaries   and  ejected   single   stars
\citep[e.g.][]{Harrington72}.    Non-hierarchical  triples   are  thus
expected to exist only at  very young ages.  The T~Tau triple HD~34700
found by \citet{Sterzik2005} could be such a case.  On the other hand,
dynamically   stable   triples    can   appear   in   non-hierarchical
configuration  in projection  on the  sky. Triple  systems  with small
period  ratios  are  more  likely  to  be  found  in  non-hierarchical
projected configurations, as HIP~54611. Here we discuss this and other
two such triple systems observed with HRCam.

\begin{figure}[ht]
\epsscale{1.0}
\plotone{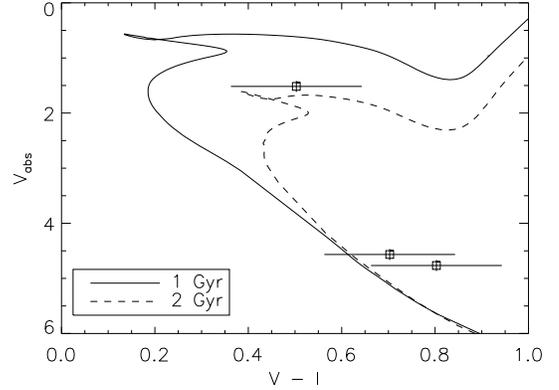}
\caption{\label{fig:CMD} Position of the three components of HIP~54611
  on the Dartmouth isochrones \citep{Dotter2008} for  solar metallicity and ages of 1\,Gyr
  and 2\,Gyr. }
\end{figure}

{\it HIP 54611.}
The spectral  type of HIP~54611 is  F5IV/V, its parallax  is $6.80 \pm
0.6$\,mas. The combined photometry in the HIP2 catalog is $V=7.24$ and
$I_C = 6.71$ mag. Assuming that our differential photometry in the $y$ and
$I$  filters also  applies  to the  $V$  and $I_C$  bands, we  compute
individual magnitudes  of A,B,C as  $V=(7.35, 10.4, 10.6)$ and  $I_C =
(6.85, 9.7, 9.8)$ mag, with 0.1  mag estimated errors. All stars appear to
be above the main  sequence (Figure~\ref{fig:CMD}), suggesting the age
of A around 2\,Gyr. It is unlikely that these stars are still evolving
towards the main sequence.

Most probably the  subsystem BC is  projected close to  the main
component A. Our photometry gives $V_{\rm BC} - V_{\rm A} = 2.39$ mag,
while  {\it Hipparcos}  measures $\Delta  Hp =  2.24  \pm 0.117$\,mag,
taking  BC as  an  unresolved  star.  The  relative  position of  A,BC
measured  by {\it Hipparcos}  also matches  this assumption.   In 1879
A,BC  was  at  $(135^\circ,  1\farcs0)$, in  1991.25  at  $(118^\circ,
0\farcs311)$, and in 2015.17  at $(115\fdg6, 0\farcs156)$. The angular
separation follows the linear  trend $\rho_{\rm A,BC}(t) \approx 0.257
- 0.0072 \; (t - 2000)$. The speed of relative motion of
7.2\,mas~yr$^{-1}$ corresponds to 1.06\,AU~yr$^{-1}$.

With projected separations of 1\arcsec  and 0\farcs2 for A,BC and BC,
respectively, and the masses  of 1.6, 1, and 1 solar for  A, B, and C,
estimated from the isochrones, we obtain the statistical periods $P^*$
of 940\,yr for A,BC and  110\,yr for BC.  The 1\arcsec ~separation and
speed of A,BC also correspond to a circular orbit with a 1-kyr period.
The  outer orbit  cannot  have a large  eccentricity,  as the  estimated
period ratio is on the order of 10, rather close to the critical ratio
of 4.7  for a circular outer  orbit. 

\begin{figure}
\epsscale{1.0}
\plotone{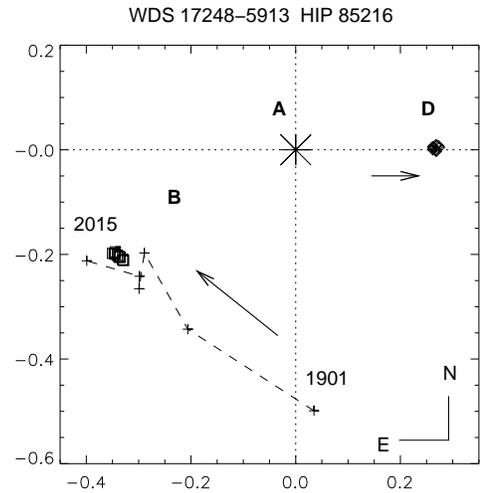}
\caption{\label{fig:I385}     Relative     motion    of     apparently
  non-hierarchical system HIP~85216 (I 385 AB and WSI~87 AD).  Crosses
  mark measures of AD,B,  squares are accurate speckle measurements of
  A,B and A,D. The scale  is in arcseconds, arrows show the
  direction of motion.  }
\end{figure}

\begin{figure}
\epsscale{1.1}
\plotone{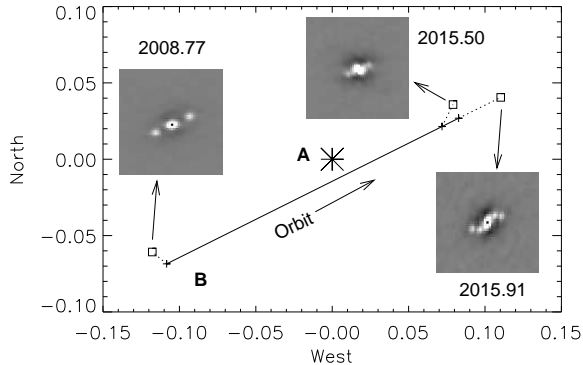}
\caption{\label{fig:STF2912} Motion of STF~2912 and its triple nature.
  The  ACFs at three  epochs are  displayed in  the inserts.  Line and
  crosses  depict the orbit of \citet{Sod1999}, the  squares denote  the measured
  positions of B (the mean of Ba and Bb in 2015.9). }
\end{figure}

{\it   HIP  85216.}   Another   ``triangular''  star   HIP~85216  (WDS
J17248$-$5913, I~385~AB  and WSI~87~AD)  was discovered with  HRCam in
2008.  All  three components are of similar  brightness, spectral type
A0V, HIP2 parallax $3.15 \pm  0.96$\,mas.  The pair AB is known since
1900.   Figure~\ref{fig:I385} shows  that B  moved since  discovery by
65$^\circ$ relative to A.   Accurate speckle measurements over 7 years
match the  general trend.  The motion  of AB can be  represented by a
tentative   orbit   with   $P=1244$\,yr,  $T_0=2374,   e=0.35,   a=0.95'',
\Omega=237^\circ,   \omega=237^\circ,   i=108.5^\circ$.   This   orbit
corresponds to a reasonable mass sum of 5.2\,${\cal M}_\odot$.

Over 7  years, the component D  moved much less than  B.  Its position
angle   is   constant,   the    separation   increases   at   a   rate
of 1.5\,mas~yr$^{-1}$.  Extrapolating back in time, the separation of AD
was 0\farcs1 in 1900 when  the AB was discovered.  With an estimated
$P^*=200$\,yr for AD we expect  a motion of 9\,mas~yr$^{-1}$, 6
times faster than actually observed. 

There are two  possibilities. Either the orbit of AD  has a very high
eccentricity and  the system is  presently seen near its  apastron, or
the actual period of AD is much  longer than the period of AB and we
see  it close  to  A only  in  projection.  As  the  orbital speed  is
proportional to $P^{-1/3}$, the observed slow motion of AD suggests a
period of  $\sim$40\,kyr rather than 200\,yr implied  by the projected
separation. In such  case the semi-major axis of AB,D  would be on the
order of 7\arcsec.  However, there  is another physical component C at
17\arcsec, threatening  the dynamical  stability of AB,D  if it  had a
7\arcsec ~axis.  The odds that  a 7\arcsec ~binary is seen at 0\farcs27
are about  $10^{-3}$, so statistically it  is more likely  that AD is
located  inside  AB,  rather  than  outside.   Considering  the  long
periods, the issue  will be hard to settle  by further observations in
the coming decades.

Whatever  is the true  organization of  the HIP~85216  multiple system,
 it  is clear  that  the apparent  non-hierarchical
configuration could be a result of projection. However, the true ratio
of separations or periods is not  very large, so the system as a whole
could be close to the limit of dynamical stability.

{\it HIP 111062.}
Figure~\ref{fig:STF2912}    illustrates    yet   another    apparently
non-hierarchical  configuration found  unexpectedly  in the  otherwise
``boring''   classical  binary   STF~2912  (HIP~111062, WDS J22300+0426).    Its  first
observation with  HRCam in  2008 shows a  simple binary, with  a minor
deviation    from   the    orbit   of \citet{Sod1999},    $P=125$\,yr,   $e=0.5$,
$i=89^\circ$.  The  orbit is  seen  nearly  edge-on.  The system  was
observed again  in 2015.5 after  passing through the  conjunction. The
ACF  looked somewhat strange,  but it  was processed  as a  binary and
showed a reasonable  agreement with the orbit. However,  in 2015.9 the
secondary was clearly  resolved into a pair of nearly  equal stars Ba and
Bb. We  plotted in Figure~\ref{fig:STF2912}  the average position  of Ba
and Bb for this epoch.

The  magnitude differences  in the  $y$  band measured  in 2015.9  are
$\Delta m_{\rm A,Ba} = 2.2$ and  $\Delta m_{\rm Ba,Bb} = 0.8$ mag. The
combined light of  Ba and Bb then leads to $\Delta  m_{\rm AB} = 1.78$
mag, in  good agreement with  $\Delta m =  1.89$ mag measured  by {\it
  Hipparcos}. This suggests that the stars Ba and Bb form a close pair
identified with the ``historic'' companion B. The projected separation
of the  pair Ba,Bb  in 2015.9 was  60\,mas, or 3.1\,AU.   Assuming the
mass sum  of Ba+Bb to  be 2 ${\cal  M}_\odot$, we estimate  its period
$P^*  \sim   3.4$\,yr.   Rapid  motion  of  this   subsystem  is  thus
expected. Considering  that the orbit of  AB is seen  edge-on and that
the pair  Ba,Bb is oriented roughly  along its plane,  we suggest that
the  two  sub-systems might  actually  be  almost co-planar.   Further
observations will quickly prove if this is indeed the case.
\citet{Trilling2007} found  IR excess  in this  and  other binary
systems, suggesting the presence  of dust. They noted that  the main star A
is evolved, above the main sequence.

The  growing  family  of  multiple  systems  with  ``weak''  hierarchy
(i.e. small ratio  of outer to inner periods)  is interesting from a
dynamical perspective,  as the motions  in the outer and  inner orbits
are  coupled  and their  representation  by  two  Keplerian orbits  is
inaccurate. The  first two systems  discussed here move slowly  on the
human  time  scale.  On  the  other  hand,  STF~2912 and  other  tight
multiples among nearby dwarfs discovered  at SOAR are much faster, and
their monitoring will yield interesting  results in the not so distant
future.

\subsection{The quadruple system ADS 9323 (HIP 71510)}

\begin{figure}
\epsscale{1.1}
\plotone{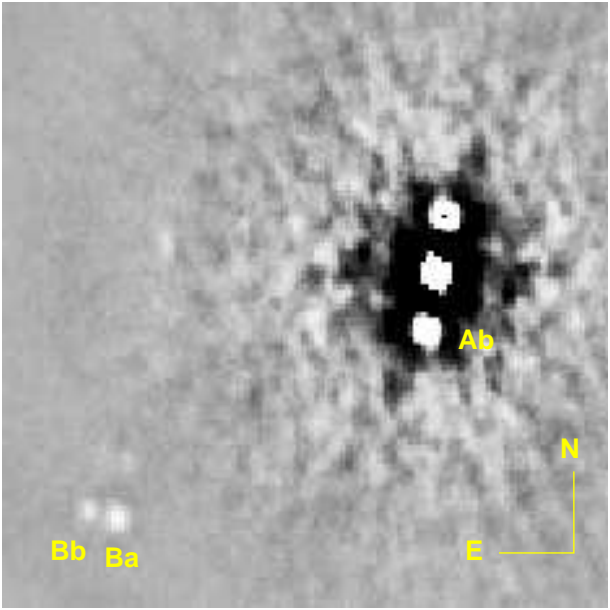}
\caption{\label{fig:9323}  Fragment of ACF  of  HIP  71510  in the  $y$  filter  
  recorded on 2009.26.  Very faint peaks corresponding  to correlation of
  Ba and  Bb with Ab are barely  seen. The
  faint details at 85\degr ~and 265\degr ~are filter ghosts.  }
\end{figure}

\begin{figure}
\epsscale{1.1}
\plotone{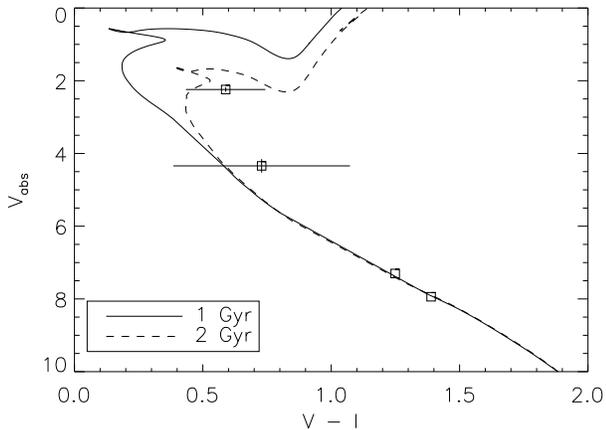}
\caption{\label{fig:iso9323}  Location  of   the  four  components  of
  HIP~71510 on  the Dartmouth isochrones  \citep{Dotter2008} for solar
  metallicity. The error bars are shown only for Aa and Ab.  }
\end{figure}

\begin{deluxetable}{ l l r r c } 
\tabletypesize{\scriptsize}    
\tablenum{5}
\tablecaption{Crude measurements of HIP~71510~B{\rm a},B{\rm b}
\label{tab:9323B} }                    
\tablewidth{0pt}     
\tablehead{
\colhead{Date}  & 
\colhead{Filt.}  & 
\colhead{$\theta$} & 
\colhead{$\rho$} & 
\colhead{$\Delta m$} \\
(yr) &   &  \colhead{ (deg)} & \colhead{ ($''$) } &   \colhead{ (mag) }
}
\startdata
2009.6290 &y &77.0 &0.100 & 1.02 \\
2014.3030 &y &73.3 &0.113 & 0.62 \\ 
2015.1716 &y &72.6 &0.103 & 0.64 \\ 
2015.4966 &I &70.1 &0.097 & 0.45 \\ 
2015.4967 &y &69.8 &0.099 & ~0.66 
\enddata
\end{deluxetable}

Observations at  SOAR revealed HIP~71510  (HD~128563, WDS J14375+0217,
ADS~9323)  as  a resolved  quadruple  system.   Its  HIP2 parallax  is
13.37$\pm$1.1 mas,  the PM is  $(-38, -70)$\,mas~yr$^{-1}$.  It  is an
X-ray  source  RX  J14375+0216.   The  outer  pair  AB  (A~2227)  was
discovered  by  R.~Aitken  in   1910  at  (138\degr,  2\farcs1).   The
secondary component B  is 4 mag fainter than A, so  the binary was not
resolved by  {\it Hipparcos}.  Presently AB closed  down to (124\fdg5,
1\farcs25).  The speed  of the relative motion of AB  in 100 years is
about 8\,mas~yr$^{-1}$,  matching its  expected orbital motion  and an
order  of magnitude less  than the  PM of  A. The  system AB  is thus
definitely physical. Its  estimated orbital period is on  the order of
500\,yr.

The  inner subsystem Aa,Ab   known  as CHR~42 was   resolved in
1984.  Its first  orbit by  \citet{Hrt2010} was  recently  updated in SOAR14:
orbital  period 21.54\,yr,  axis 0\farcs143,  eccentricity  0.844.  The
HIP2 parallax and orbital elements give the mass sum of 2.64 ${\cal M}_\odot$.

The  system was  first  observed  at SOAR  in  2009.26. The  secondary
component B was resolved into a 0\farcs1 pair (Figure~\ref{fig:9323}).
However, the  data processing  tools do not  provide for  fitting four
stars,  and  the  binarity  of  B  was  not  mentioned  explicitly  in
TMH10. The speckle observations in  2014 and 2015 were made mostly for
the purpose of  following the orbit of Aa,Ab,  which closed down.  The
subsystem Ba,Bb  was noted,  but still not  measured.  In  2015.46 the
Aa,Ab pair  closed down  almost below the  diffraction limit,  and the
data were processed as a triple system (A,Ba,Bb) for the first time.

The Ba,Bb pair was  measured approximately 
with an accuracy of $\sim$0.5  pixel (8\,mas).  The results are listed
in Table~5.  The pair  is in a slow retrograde motion.  Its
period estimated from separation is on the order of 20\,yr. The motion
of AB is also retrograde, while Aa,Ab is in direct motion.

SIMBAD does not  provide the $BVRI$ photometry of the  star, but it is
found  in  \citet{Metanomsky1998}.   The  differential  photometry  is
furnished  by  the  SOAR   speckle.   The  results  are  assembled  in
Table~6. The  faintness of Ba and  Bb and approximate
data processing  make its  speckle photometry quite  uncertain, while
the magnitudes of Aa and Ab are established securely.

\begin{deluxetable}{ l l r r r r } 
\tabletypesize{\scriptsize}    
\tablecaption{Photometry of HIP 71510
\label{tab:71510ptm} }                    
\tablenum{6}
\tablewidth{0pt}     
\tablehead{
\colhead{Component}  & 
\colhead{$V$ or $y$}  & 
\colhead{$I_C$} & 
\colhead{$J$} & 
\colhead{$K_s$} \\
\colhead{combination} & 
\colhead{(mag)} & 
\colhead{(mag)} & 
\colhead{(mag)} & 
\colhead{(mag)} 
}
\startdata
A+B       &  6.45  &   5.83  & 5.45     & 5.11     \\
A$-$Aa    &  2.10  &   1.96  & \ldots   &  \ldots \\
Ba$-$Aa   &  5.06  &   4.4*  &\ldots     &  \ldots \\
Bb$-$Ba   &  0.64  &   0.5*  &\ldots     &  \ldots 
\enddata
\end{deluxetable}

The combined light of the system  is dominated by the component Aa. It
is classified as  F9V, which matches the combined  color. However, the
absolute  $V$ magnitude  of Aa  is 2.2 mag,  about 2  mag brighter  than a
normal F9V  dwarf. We  cannot question the  HIP2 parallax (it  gives a
reasonable  mass sum of  Aa,Ab), so  Aa is  definitely above  the main
sequence.  Treating  the system  as a single  star, \citet{Casagrande2011}
 derived $T_e = 6343$\,K, mass 1.75 ${\cal M}_\odot$, and age about 1.5\,Gyr.

Figure~\ref{fig:iso9323}  shows  the  position  of the  components  of
ADS~9323   on  the   Dartmouth   isochrones  \citep{Dotter2008}.   The
differential magnitudes of  Ba and Bb in the  $I_C$ band were adjusted
by trial  and error so  that they fall  on the main  sequence (numbers
with asterisks in  Table~6); they do not contradict  the crude speckle
photometry.  The system is slightly older than 2\,Gyr.  The mass of Aa
from isochrones is about 1.62 ${\cal M}_\odot$, the mass of Ab is 1.22
${\cal  M}_\odot$,  the  mass  sum   of  Aa+Ab  is  then  2.82  ${\cal
  M}_\odot$. It matches the mass derived from the orbit, 2.64$\pm$0.65
${\cal  M}_\odot$,  within its  error  (which  is  mostly due  to  the
parallax uncertainty). The  masses of Ba and Bb  are estimated as 0.74
and 0.68 ${\cal M}_\odot$.

\section{Summary}
\label{sec:concl}

The extremely productive speckle  interferometry program at SOAR gives
a unique look on the structure and dynamics of multiple systems in the
solar  neighborhood.   We  focus  on  close  pairs   resolved  by  our
predecessors, by {\it Hipparcos}, and  in the previous speckle runs at
SOAR.  The inventory of new  pairs is enlarged by 27 first resolutions
reported here.

The 1303  measurements made at SOAR  in 2015 are  used for calculation
and improvement  of orbits, both by  our team (e.g. the  198 orbits in
SOAR2014, where some measurements  reported here were already used) and
by  other  authors, e.g.  WDS  J22504$-$1744 \citep{Docobo2015}.   Slow
motion of  some binaries means that  this material will be  used for a
long time.   Its value will increase further  when accurate parallaxes
from {\it  Gaia} become available.   The short duration of  this space
mission makes long-term monitoring  of orbital motion from the ground,
such as provided here, its essential complement.

Although measurement of stellar masses  is still needed in many areas,
orbital  elements give  a  wealth of  other   information.   For
example,  their statistics  are   useful for  testing  theories of
binary  formation.  This is  particularly true  for systems  with more
than  two bodies,  e.g.  hierarchical  multiples and  binaries hosting
planets.   Period ratios, eccentricities,  and angles  between orbital
planes allow dynamical analysis of these complex astrophysical systems
to gain insights on their  origin and evolution.  We characterize here
motions in  three apparently  non-hierarchical triple systems  and one
quadruple discovered with HRCam. 

High angular  resolution and deep  dynamic range allow us  to discover
new   components  in  previously   known  binaries,   sometimes  quite
unexpectedly   (Figure~\ref{fig:ACF}).   Here   we   established  that
$\epsilon$~Cha consists  of three  nearly equal stars,  not of  two as
believed  before.  This  is  the  most massive  member  of  a  young
association  in the  solar  vicinity.  The period  of  the inner  pair
discovered here  is only a few  years, so its  further monitoring will
soon provide masses of B9V pre-main-sequence stars in a triple system.

\acknowledgments 

We thank the operators of SOAR D.~Maturana, P.~Ugarte, S.~Pizarro, and
J.~Espinoza for efficient  support of our program and  we are grateful
to the  SOAR Director J.~Elias for granting  technical time
for this program.

R.A.M. acknowledges  support from the Chilean Centro  de Excelencia en
Astrof\'{i}sica y Tecnolog\'{i}as Afines  (CATA) BASAL PFB/06, and the Project
IC120009 Millennium Institute of  Astrophysics (MAS) of the Iniciativa
Cient\'{i}fica Milenio  del Ministerio de  Econom\'{i}a, Fomento y  Turismo de
Chile.

This work  used the  SIMBAD service operated  by Centre  des Donn\'ees
Stellaires  (Strasbourg, France),  bibliographic  references from  the
Astrophysics Data  System maintained  by SAO/NASA, and  the Washington
Double Star Catalog maintained at USNO.

{\it Facilities:}  \facility{SOAR}.



\clearpage


\end{document}